\documentclass{aa}
\usepackage{graphicx}
\usepackage{txfonts}
\usepackage{siunitx}
\usepackage{natbib}
\begin{document}

   \title{Incorporating Astrochemistry into Molecular Line Modelling via Emulation}

   \author{D. de Mijolla
          \inst{1}
          \and
          S. Viti\inst{1}
          \and
          J. Holdship\inst{1}
          \and
          I. Manolopoulou \inst{2}
          \and
          J. Yates \inst{1}\fnmsep
          }

   \institute{Department of Physics and Astronomy, University College London, Gower Street, WC1E 6BT, UK
         \and
Department of Statistical Science, University College London, Gower Street, WC1E 6BT, UK\\
             }

   \date{Received xx xx,xx; accepted xx x, xx}

 
  \abstract{In studies of the interstellar medium in galaxies, radiative transfer models of molecular emission are useful for relating molecular line observations back to the physical conditions of the gas they trace. However, doing this requires solving a highly degenerate inverse problem. In order to alleviate these degeneracies, the abundances derived from astrochemical models can be converted into column densities and fed into radiative transfer models. This enforces that the molecular gas composition used by the radiative transfer models be chemically realistic. However, because of the complexity and long running time of astrochemical models, it can be difficult to incorporate chemical models into the radiative transfer framework. In this paper, we introduce a statistical emulator of the UCLCHEM astrochemical model, built using neural networks. We then illustrate, through examples of parameter estimations, how such an emulator can be applied on real and synthetic observations.}

   \keywords{astrochemistry --
                radiative transfer --
                methods: statistical --
                ISM: Molecules --
                galaxies: abundances
               }

   \maketitle
%

\section{Introduction}

Molecules form in the interstellar medium provided it is dense enough for collisions to bring the chemical reactants together and cool enough to suppress the complete dissociation of chemical products. Observations of the interstellar medium in and outside of our galaxy have revealed a rich and diverse chemistry \citep{Shematovich2012}, spanning across a wide range of physical environments. As stars form through the collapse of over-densities inside this optically opaque dense interstellar medium \citep{Young1991}, the study of molecular gas can provide a window into the star-formation process and shed light onto the lifecycle of galaxies.

Typically, the chemistry of the cold and dense interstellar medium is probed through measurements of molecular lines. In order to interpret these, radiative transfer models are used to relate the line strengths back to the physical conditions of the interstellar medium they trace. RADEX (see \cite{VanderTak2007}) is a popular non-local thermodynamic equilibrium radiative transfer model that models, for a given set of physical conditions, the expected strength of  molecular lines. By matching the observed molecular lines to predictions from grids of such models, it is possible to constrain the density, temperature and column density of the molecular gas (e.g. \cite{Viti2017} \cite{Tunnard2016}, \cite{Salak2018}). 

However, as molecular line intensities are dependent on a complex interplay between the physics, chemistry and radiative-transfer of the observed region, their interpretation is often ambiguous. As such, estimating chemical parameters via the use of radiative transfer models is a notoriously degenerate problem. Even under a set of idealized assumptions, when one assumes that the gas can be accurately represented using a single component at a unique temperature and density, there exist wide ranges of parameter values capable of fitting a set of observations (see \cite{Tunnard2016}, \cite{Kamenetzky2018}). These degeneracies are amplified when studying external galaxies as the telescope beam sizes often encompass a wide variety of different physical environments that must be disentangled.

Since line intensity predictions obtained from radiative transfer models depend on the column densities of the studied region, it is customary to treat column densities as free parameters of the radiative transfer modelling to be constrained alongside the temperature and density. Unfortunately, this makes the radiative transfer modelling highly degenerate, oftentimes leading to many very different models being able to fit the same observations. There have been attempts to address these degeneracies through the use of astrochemical models. 

Astrochemical models, such as UCLCHEM presented in \cite{Holdship2017}, are computational codes designed for modelling the chemical composition of gas under well-defined physical conditions. This is done through the numerical integration of a set of differential equations constructed from a network of chemical reactions. It has been proposed, for example in \cite{Viti2014} and more recently in \cite{Viti2017} and \cite{Harada2018} to convert the outputs of chemical models into column densities to be used as inputs to radiative-transfer models. Without including chemistry into the forward model, the parameter retrieval can give retrievals that are inconsistent with our chemical knowledge as the forward model has no knowledge of which species should be abundant for given conditions. With inclusion of the chemistry, not only are column densities no longer free parameters, leading to tighter bounds on the retrieved parameters, it becomes possible to constrain the parameters driving the chemistry, such as for example the metallicity and cosmic-ray ionization rate. However, the widespread integration of astrochemical models into the radiative transfer process is hindered by their long running-times and their complexity. The long running times result in even relatively small grid of astrochemical models requiring a large amount of computational resources.

In this project, we address both of these issues through the creation of a statistical emulator for the UCLCHEM astrochemical model. The statistical emulator is built using a set of neural networks trained to find a multidimensional fit to a training dataset of chemical simulations. The emulator offers a considerable speed-up in modelling time, being able to estimate molecular abundances in milliseconds, much faster than full chemical models which run in minutes. On top of the speed-up, the UCLCHEM emulator has been simplified, having dependencies on only six variables; the density, temperature, metallicity, visual extinction, cosmic-ray rate and radiation field rate of the region being modelled. With our emulator, the complexity and computational power required to include chemical models into the radiative transfer process, has been considerably reduced. As a by-product, we have also created a RADEX emulator for a selected few key species for an even faster inference.

Although emulators have gained some traction in the cosmology community (e.g. \cite{Schmit2018} and \cite{Kwan2013}), they remain uncommon in astrochemistry. The astrochemical community instead usually favours comparison to tables of precomputed models ( e.g. \cite{Mondal2018}, \cite{Meijerink2007}, \cite{Maffucci2018b}, \cite{Bisbas2019}). In \cite{Grassi2011}, a neural network was trained to replace the chemical network calls in N-body simulations. Our work differs in that, by modelling the full chemical evolution starting from a limited set of realistic initial conditions, we have restricted the parameter space. This avoids issues of error propagation over each time-step and allows for using more complex chemical models.

In section \ref{Astronomial_models}, we give some details on the physical and chemical models being emulated as well as the emulation procedure. In section \ref{Uclchem_emulator} and \ref{Radex_emulator}, we give a technical overview of our emulation procedure and quantify, over our selected parameter range, our emulator's ability to accurately predict molecular abundances and intensities. In section \ref{Bayesian_retrieval}, using a set of toy observations, we demonstrate how the emulator can help to lift some of the degeneracies present in radiative transfer modelling. In section \ref{Applications_Real}, we further apply the emulators to ALMA observations of the nearby prototypical Seyfert 2 galaxy NGC1068 (presented in \cite{Garcia-Burillo2014} and \cite{Viti2014}) and touch upon some of the weaknesses and strengths of our emulator.


\section{Modelling molecular gas} \label{Astronomial_models}

In this section we give a brief overview of the specifics of the chemical and radiative transfer models used for emulation. We explain how we combined these to create a simple forward model capable of reproducing observations of the interstellar medium. We finish the section by covering how we can use a neural network to create an emulator of astronomical models.

\subsection{Chemical Models}

UCLCHEM is a time-dependent gas-grain open-source chemical model described in \cite{Holdship2017}. In UCLCHEM, chemical evolution of the gas is divided into two phases. In the first phase (phase I), supposed to approximate the molecular gas formation processes, the gas starts in a diffuse atomic state and evolves following a freefall collapse. In the second phase of the model (phase II), the physical conditions are modified so as to approximate specific observable environments. For an extragalactic application, this could involve high cosmic-ray rates as would be expected in an AGN dominated galaxy, or high-uv flux as would be expected in a starburst galaxy.

For our models, during phase I, the gas started at a density of $100$ \si{cm^{-3}}, and was then compressed, via freefall, to a final density, left as a free parameter. This freefall collapse was isothermal, with a gas temperature of 10K.  The radius of the region, which was also used to calculate the visual extinction, was allowed to vary. During this phase, gas phase desorption and freeze-out on the dust grains proceeds.

During phase II, the gas was assumed to have reached its final density and the physical parameters were varied so as to model a range of environments. The models were all run for $10^7$ years, long enough for the gas to reach chemical equilibrium.  The parameters that were allowed to vary were:

\begin{itemize}
  \item The temperature (T): The temperatures of the phase II models were increased over time up to a value T following the same procedure as in \cite{Viti2004}, where the temperature increases with time as a function of the luminosity of an evolving star. Of course the temperature dependence on time will differ depending on what objects the chemical model is simulating but for the purpose of this study we simply adopted the procedure already present in UCLCHEM. The models were then further run at fixed temperatures until a cumulative time of $10^7$ years.
  \item The gas density (n): The phase I models  were run following a parametric freefall collapse \citep{Rawlings1992} until a density n was reached. The density was then kept constant during phase II.
  \item The metallicity (m$_Z$): The initial atomic abundances used by the chemical network were constrained to be a fraction of the solar metallicity. We have defined metallicity as a multiplicative factor with a metallicity of 1 corresponding to elemental abundances as found in \cite{Asplund2009a}.
  \item The cosmic-ray ionization rate ($\zeta$): This parameter represented the cosmic-ray ionization rate used in Phase II. Additionally, as we do not model the X-ray ionization rate, we use the cosmic-ray ionization rate as its proxy \citep{Xu2016}. As noted in \cite{Viti2014} this approximation has its limitations in that X-ray heating is more efficient than cosmic-ray heating.
  \item The uv-photoionization rate ($\chi$): This parameter represented the uv-photoionization rate used for phase II models. The uv-photoionization is measured in Draine where 1 Draine is equivalent to 1.6$\times$10$^{-3}$ erg/s/cm$^2$ \citep{Draine1978,Draine1996}
  \item The visual extinction ($A_{\rm V}$): In UCLCHEM the visual extinction of the molecular gas is controlled by the size of the modelled region.
\end{itemize}

At its core, the UCLCHEM chemical model is centred around a chemical network specified by the user. For this project we used a chemical network based on the UMIST database \citep{McElroy2013}. For each time-step, a set of coupled ordinary differential equations is generated from the chemical network and solved. We refer the reader to the UCLCHEM release paper \citep{Holdship2017} for a thorough overview of the effect of various parameters on these rate equations.

\subsection{The Radiative Transfer Model}

The non-LTE radiative transfer code RADEX \citep{VanderTak2007} in conjunction with collision files obtained from the LAMBDA database \citep{Schoier2005} was used for estimating line intensities strengths. RADEX is a non-LTE radiative transfer model that decouples the non-local radiation field from the local level population calculation through the escape probability approximation.

For our emulator, all radiative transfer models were run with H$_2$ as the unique collisional partner, assuming a background temperature of 2.7K and assuming a spherical geometry. As both \cite{Krips2011} and \cite{Viti2014} found that  different geometry choices in RADEX gave comparable outputs for the fitting of CO, HCN and HCO$^+$ in a nearby galaxy we restrict ourselves to using a spherical geometry.

\subsection{The Forward-Model \label{forward}}

Using molecular line intensities to constrain the physical conditions of the interstellar medium is an inverse problem. In practice, this can be tackled by comparing synthetic line intensity predictions obtained using a forward model with the measured molecular lines. In this paper we contrast two distinct forward modelling approaches.

\begin{itemize}
\item The forward model can encompass solely the radiative transfer physics (from now on referred to as "chemistry-independent"). This is the more established methodology for analysing molecular lines (\cite{Imanishi2018}, \cite{Michiyama2018}).
\item The forward model can encompass the chemistry of the molecular gas in addition to the radiative transfer (from now on referred to as "chemistry-dependent"). This approach is less common but has been used in \cite{Harada2018}, \cite{Viti2017} and \cite{Viti2014}. 
\end{itemize}

In the chemistry-independent forward modelling approach, the temperature, gas density, line-width, column-densities and beam-filling factor are the only parameters allowed to vary. The first four parameters correspond to the input parameters used in the RADEX modelling. The beam-filling factor is treated as a multiplicative scaling factor.

In the chemistry-dependent forward model, the parameters allowed to vary are the inputs to the chemical model, the line widths and the beam-filling factor. In this case the synthetic observations are created by converting the output abundances from the chemical model into column densities and using these, as well as the final chemical model temperature and density, as inputs to the radiative transfer model. The predicted intensities from the radiative transfer model are then transformed into mock observations by multiplying them by a beam-filling factor. In order to convert abundances into column densities, the fractional abundances are multiplied by the column densities of hydrogen as measured at 1 mag and by the visual extinction. We used $N(\rm H_2)=1.6\times10^{21} \rm cm^{-3}$  for the column density of hydrogen at 1 mag. This conversion procedure is known as the "on-the-spot" approximation (e.g. \cite{Dyson1997}).  

These models can easily be extended to include more than one phase of gas. For example for a two-phase gas, one can run two single-phase models and add up their intensities (after rescaling by the beam filling factor). A summary of the newly introduced  parameters and their notation is as follows:
\begin{itemize}
  \item The linewidth ($\Delta v$): This is the linewidth used as an input in the RADEX radiative-transfer calculations. For simplicity, we assume in our forward model that all molecular lines share a common linewidth.
  \item The filling factor (f): For each phase, the output intensities from RADEX are rescaled by a beam filling factor representing the fraction of the beam occupied by the emission.

\end{itemize}

\subsection{Artificial Neural Networks}

Artificial Neural Networks (ANN) are a class of algorithms used for learning mappings between an input space and an output space \citep{Goodfellow:2016:DL:3086952}. ANN are trained by tuning a set of parameters to match a training dataset composed of input-output pairs. The building blocks of ANN are nodes, called neurons, connected together by edges. In an ANN, information is passed between nodes through these edges and combined through non-linear functions to obtain a mapping from the input to the output space.

\begin{figure}
	\includegraphics[trim=0 0 300 0, clip,width=\columnwidth]{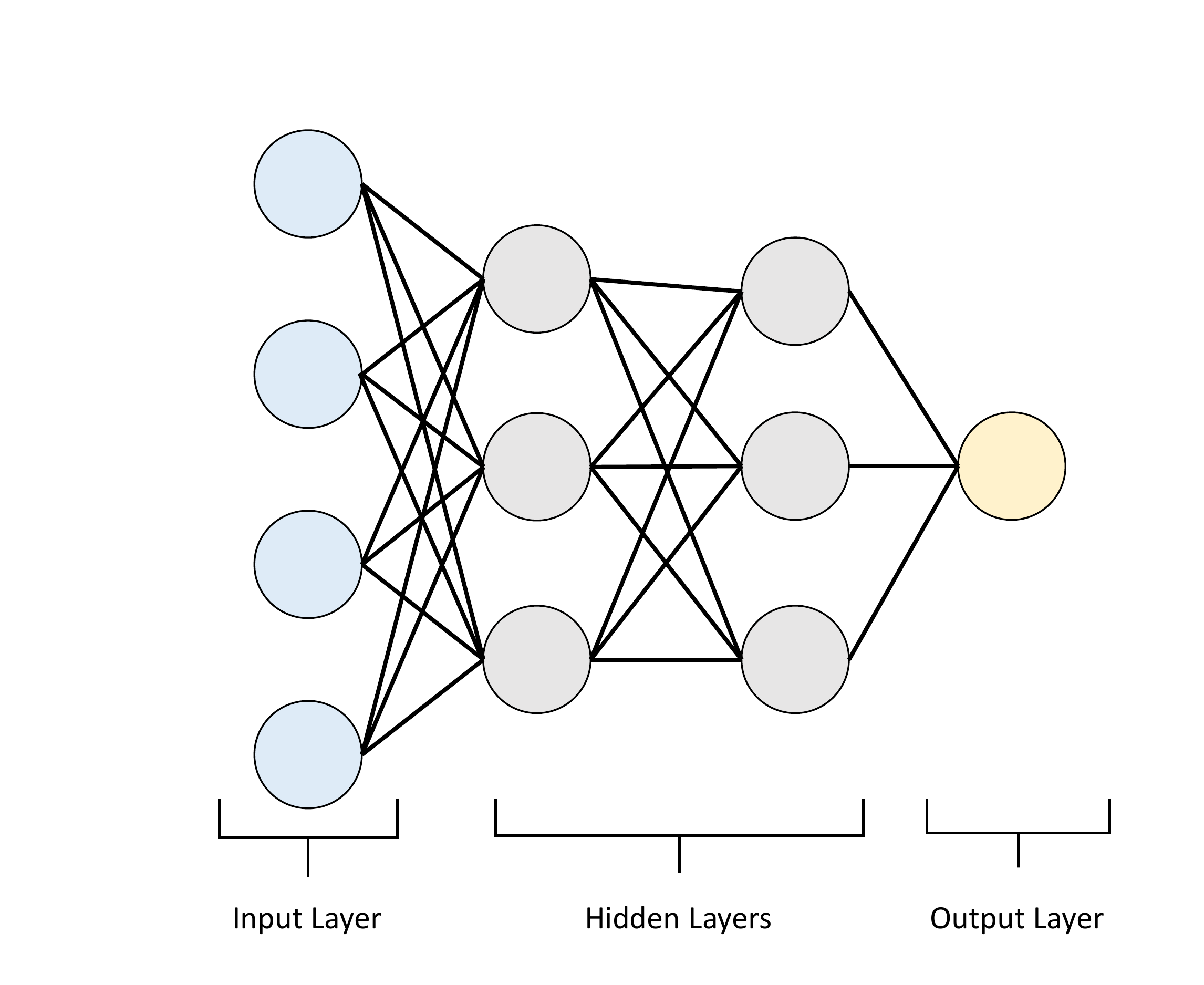}
    \caption{Illustration of a multilayer perceptron neural network.}
    \label{fig:feedforward neural network}
\end{figure}

In feed-forward neural networks (Figure \ref{fig:feedforward neural network}), as used in this project, the neurons are organized into sequential layers. The neurons from each layer are connected by edges to every neuron of the succeeding and preceding layer. Neurons are place-holders in which numbers are stored with the first layer containing the inputs fed to the neural network and the last layer containing the associated predictions from the neural network. All the neurons in other layers are place-holders for intermediate values used in the calculations. The predicted outputs, given an input, are found by successively calculating for each layer the neurons values starting from the input layer up to the output layer. For neuron $j$, the formula for calculating its value $a_j$ is $a_j = \Phi\left ( \sum_{i} w_{ij} a_i +b_j \right )$ where $\sum_{i}$ refers to a sum over all the neurons of the previous layer, $b_j$ is a parameter associated to every individual neuron usually referred to as bias and $w_{ij}$ is a parameter associated to every individual edge called the weight of the edge. The function $\Phi \left (  x \right )$, often called the activation function, is a non-linear function, whose presence makes it possible to approximate non-linear combinations of the inputs. For this specific project a rectified linear unit (ReLU) activation function was used \citep{Vinod2010}.

The neural network parameters (bias and weights) are determined using a set of training data. Typically, the training data contain example inputs and their associated outputs. The neural network then attempts to fine-tune the parameters so as to minimize a user-specified loss function, designed to assess how well the neural network predictions match the example outputs.

In this project we utilize neural networks as emulators. This consists of using the inputs of a model as the inputs to a neural network and training the neural network to reproduce the outputs of the model. By training a neural network to emulate a model, it becomes possible to bypass the model. This is advantageous if the model is computationally intensive to run, as it allows for obtaining samples at a fraction of the original computational cost.

In the application presented here, because there is a strong overlap in the parameter space explored when modelling different galaxy observations, very similar grids of parameters are run even when interpreting radically different observations. This makes the use of an emulator particularly advantageous as the overhead required in training an emulator will quickly be smaller than the accumulated run-time from running redundant models. 

\section{UCLCHEM Emulator} \label{Uclchem_emulator}

In this section we discuss the creation and evaluation of our emulator.

\subsection{The Training dataset}

A dataset of $N=120000$ chemical models was generated. The parameter ranges for the training dataset can found in Table \ref{tab:parameters_introduction}. Since the emulator is only able to interpolate and not extrapolate, these parameter ranges define the usable range of the emulator.
\begin{table}
\centering
\begin{tabular}{ p{1.4cm} p{1.4cm} p{1.4cm} p{2.0cm}  }
 \hline \hline
 \multicolumn{4}{c}{Parameter sampling ranges} \\
 \hline
 Parameter& minimum & maximum & unit\\
 \hline
 $A_{\rm V}$  & $1$    & $100$&   mags \\
n&   $10^{4}$  & $10^6$   & cm$^{-3}$\\
 $\zeta$  &1 & $10^3$&  $1.3\times 10^{-17} \rm s^{-1}$\\
  $\chi$   & $1$ & $10^3$&  Draine \\
 T &   $10$  & $200$ & K \\
 m$_Z$ & 0  &2   &  Z\\
\hline
\end{tabular}
\caption{Emulator parameters and their range.}
\label{tab:parameters_introduction}
\end{table}

A Latin hypercube Sampling Scheme was used for generating the samples in the training dataset. Latin hypercube sampling \citep{McKay1979} is a statistical method for generating near random samples particularly suitable for exploring parameter spaces under a restricted computational budget. It has been used, for example, in \cite{Schmit2018} for emulation of epoch of reionization simulations and in \cite{Bower2010} for emulation of semi-analytical galaxy models.

\subsection{The Algorithm}

Feed-forward neural networks were used to predict molecular log-abundances from the UCLCHEM inputs. A separate neural network was trained for each molecular species in our chemical network with each neural network sharing the same five-layer architecture. The first layer was 6 neurons wide, with each neuron being assigned to one of the 6 input parameters. The next 3 layers were successive hidden layers of width 200, 100 and 50. Finally the last layer represented the predicted output by the neural network. All of the layers, used a ReLU activation function \citep{Vinod2010}.

For each molecular specie, the neural network was trained through the back-propagation algorithm (see \cite{Rumelhart1986}) to minimize the MSE loss over the whole dataset between the chemical simulation outputs and the neural network outputs 
\begin{equation}
    \sum_{1}^{N} (y_i-\hat{y}(x_i))^2
\end{equation}
with $y_i$ the  log10-abundance predicted by UCLCHEM, $x_i$ the neural network input parameters (see Table \ref{tab:parameters_introduction}) and $\hat{y}$ the neural network prediction. As abundances cover several order of magnitude in scale, in order to treat the whole parameter range equally, we trained the neural network to minimize the log-abundances. In addition, the input parameters were scaled to lie within the 0-1 range before being fed to the neural network by using the following transform:
\begin{equation}
    \rm X_{\rm scaled} = \frac{\rm X-\rm min(X)}{\rm max(X)-\rm min(X)},
\end{equation}
where X is an input parameter before rescaling. 

Training of the neural networks was done using the pytorch framework \citep{paszke2017automatic}. The neural network parameters were optimized using Adam \citep{DBLP:journals/corr/KingmaB14} with an initial learning rate of 0.001 and training occurring over 20 epochs of batch sizes 500.

\subsection{Error analysis}

\begin{figure*}
	\includegraphics[width=\linewidth]{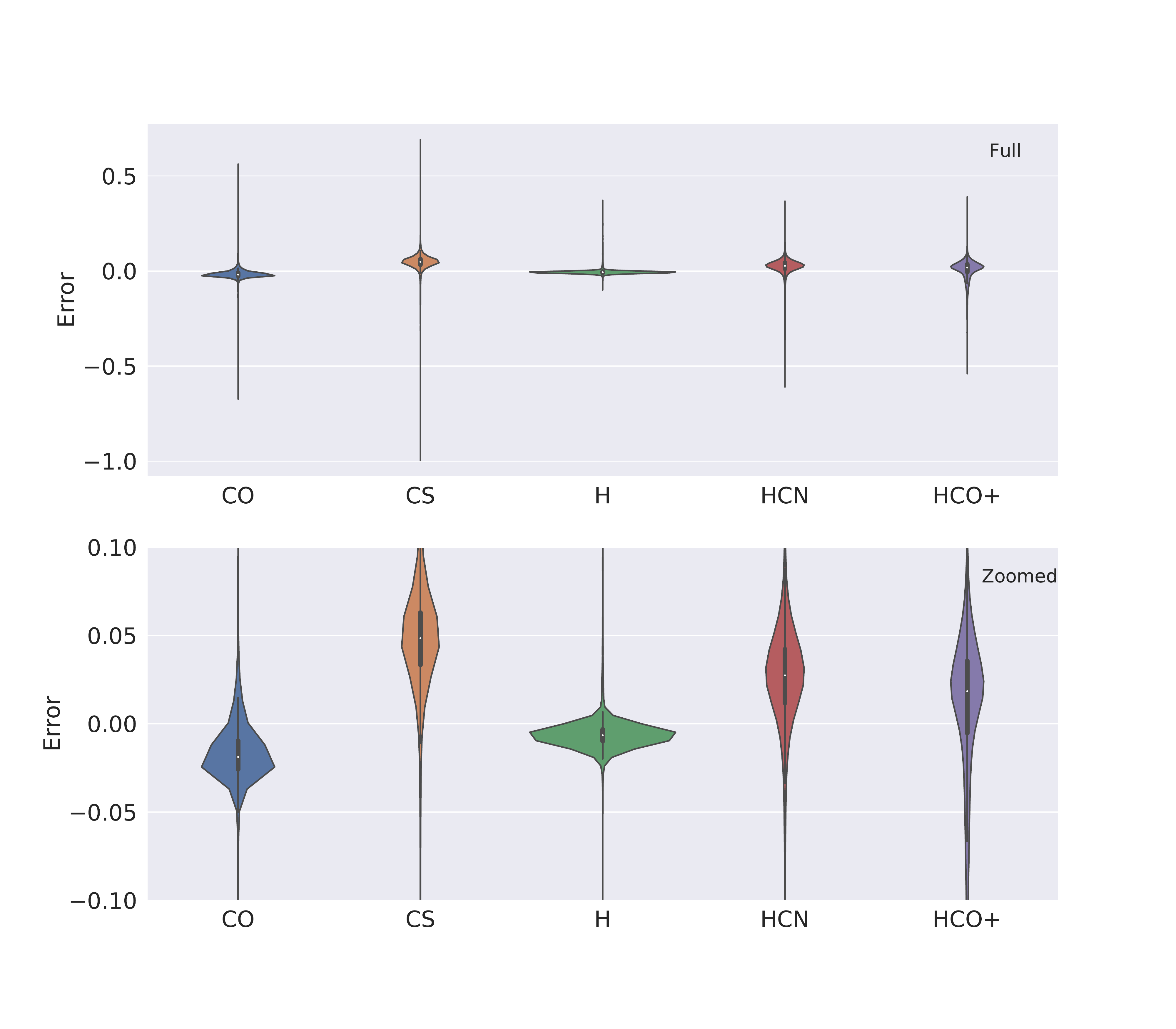}
    \caption{Violin plot of the distribution of difference between the log10 abundance predictions from the astrochemical models and from the emulator using a kernel density estimate from the 10000 simulations in the validation dataset for CO, CS, H, HCN and HCO$^+$. The bottom plot is a zoomed-in version of the top plot. In the bottom plot, the thick black lines represent the interquartile range and the thin black line the 95\% confidence intervals.}
    \label{fig:violinChem}
\end{figure*}

We quantified the approximation error in our emulation procedure by plotting the distribution of differences between the emulated log-abundances and the UCLCHEM log-abundances for a validation dataset, constructed by running 10000 UCLCHEM models with randomly sampled input parameters, (Figure \ref{fig:violinChem}). Reassuringly, the emulation error is small for all of our plotted species, with $95\%$ of the models agreeing with the emulator predictions to within a multiplicative factor of 1.13 (a log10-abundance  difference of 0.05). At the tail of the distribution, the worst emulator predictions are still within a factor of 10 of the correct abundances (a log10-abundance difference of 1).

.

\subsection{Effect of the dataset size}

\begin{figure}
	\includegraphics[width=\columnwidth]{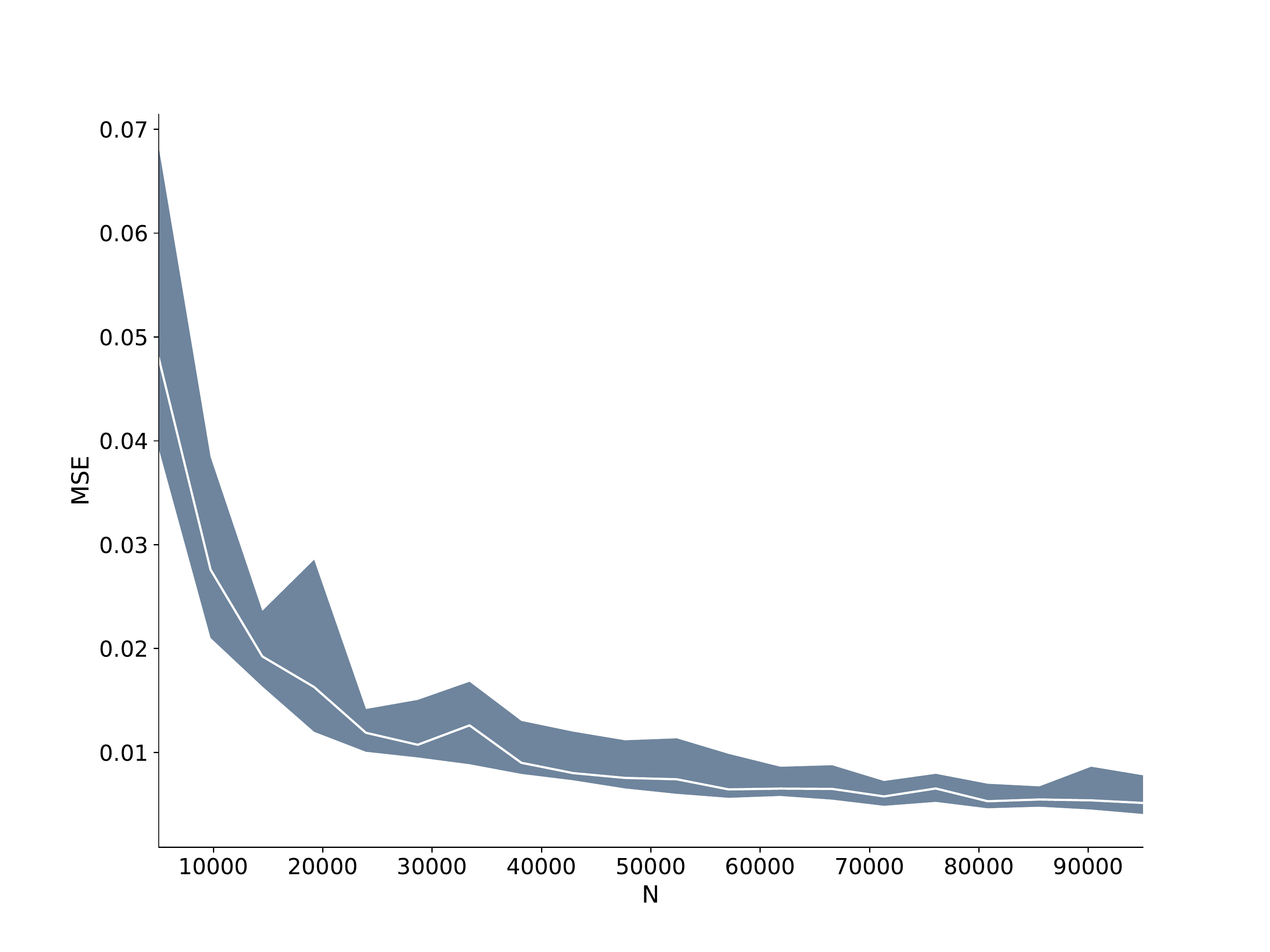}
    \caption{A plot of the effect of training set size on emulator prediction. On the y-axis, the mean squared error between the log10 ground truth abundances and neural network prediction evaluated on the validation dataset. On the x-axis, the size of the training dataset. The shaded area represents the spread of mean squared error obtained across runs with the $68.2\%$ percentiles centered around the mean being shaded.}
    \label{fig:simulationN}
\end{figure}

We investigated the effect the training dataset size had on the predictive power of our emulator. To do this, we trained our neural networks on different-sized subsets of the training dataset and quantified how the training dataset size affected predictions. For each dataset size, we averaged the validation set errors over multiple runs. The outcome is shown in Figure \ref{fig:simulationN}. From this Figure, we can see that further increasing the dataset size beyond the 120000 samples already used would only offer very marginal improvements to our emulator's predictive ability.

\section{Radiative Transfer Emulator} \label{Radex_emulator}

In addition to creating an emulator for UCLCHEM, we created a radiative transfer emulator that replicated the results obtained by RADEX for $J<10$ molecular line transitions. As each molecule required running a brand new set of RADEX models we restricted ourselves to emulating HCO$^+$, HCN, CO and CS. Although individual RADEX models are relatively quick to run, exploring a high-dimensional radiative transfer parameter space can require hundreds of thousands of simulations. This makes the use of a RADEX emulator particularly useful, especially in cases where a slight loss in accuracy is acceptable, such as when used in conjunction with chemical models, which themselves have high associated uncertainties.

\subsection{Training Dataset}

The emulator took  as inputs the temperature, density, line-width and molecular column densities. By exploiting the degeneracy between line-width and column density for optical-depth, we were able to remove the line-width dependency from our training dataset. All the parameter ranges were kept the same as for the UCLCHEM emulator, with  the maximal column densities for our RADEX emulator rounded-up to be an order of magnitude larger than the maximal column densities in our chemical model dataset. A Latin hypercube sampling scheme was run over the chosen parameter range. We sampled from the log-column density.

The escape-probability formalism that underpins RADEX can break-down for some parameter choices, leading to spurious intensities. To mitigate this effect we applied a visually chosen cut-off to the intensities in our training-set, all simulations with intensities higher than the cutoff were excluded from the dataset.

\subsection{Algorithm}

We used the same neural network preprocessing, training, and architecture as that used for the UCLCHEM emulator. We trained the neural network using an L1 loss
\begin{equation}
    \sum_{1}^{n} \left | y_i-\hat{y}(x_i) \right |
\end{equation}{}
with $ y_i$ referring to the intensity predicted by RADEX, $x_i$ the corresponding RADEX input parameters and $\hat{y}$ the neural network prediction on a dataset of approximately 100000 RADEX outputs. Because our method for removing spurious intensities was not perfect, there still remained some spurious models. This meant that an L1 loss, which put less emphasis on fitting every single data point perfectly, was more suitable then a L2 loss. To prevent the neural network from predicting non-physical values, we rounded all negative intensity predictions to zero.

\begin{figure}
	\includegraphics[width=\columnwidth]{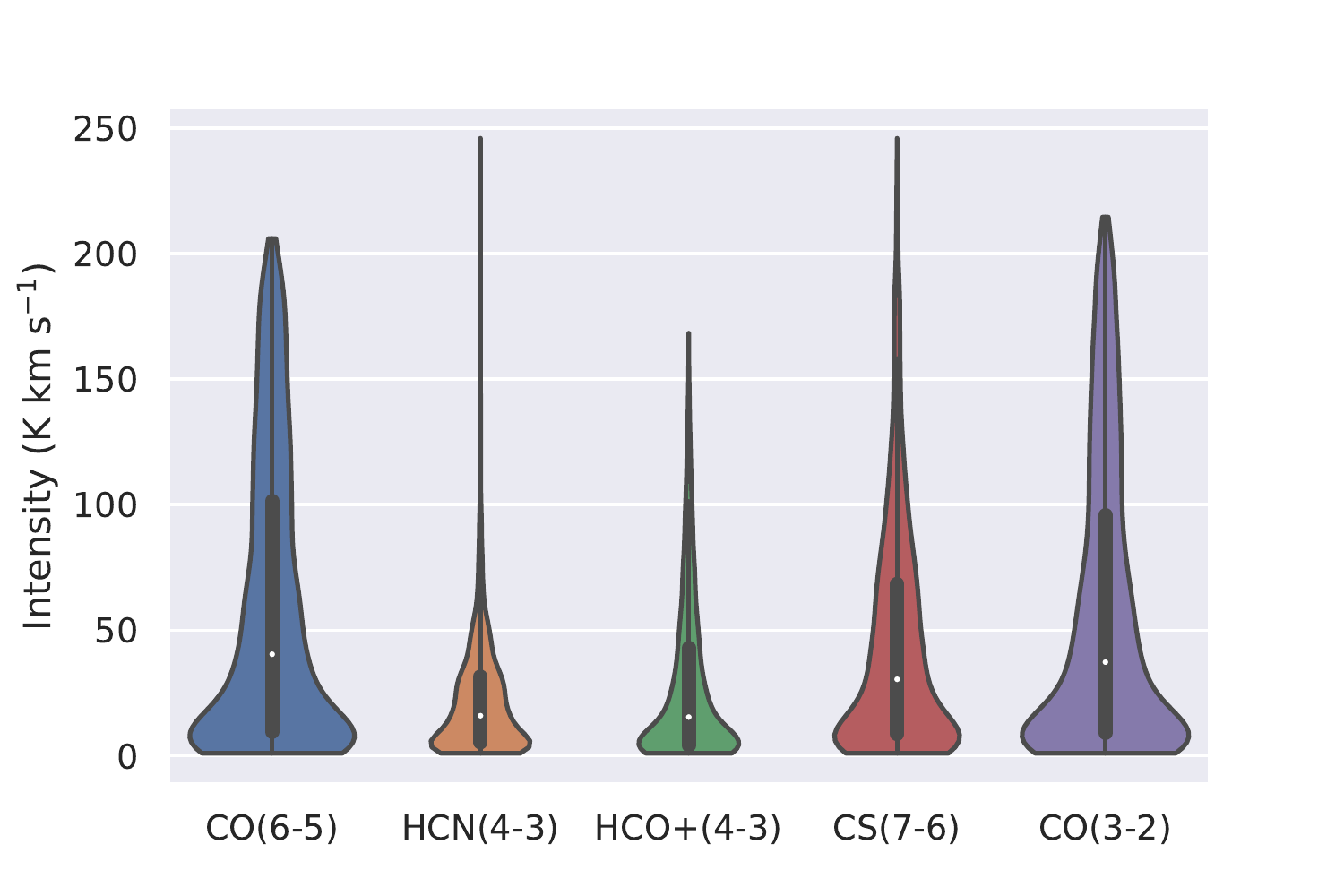}
    \caption{Violin plot of the distribution of RADEX intensities for different molecular lines. The distributions are obtained using a kernel density estimate from the 10000 simulations in the dataset. The thick black lines represent the interquartile range and the thin black lines the 95\% confidence intervals.}
    \label{fig:violinIntensityRadex}
\end{figure}

\begin{figure}
	\includegraphics[width=\columnwidth]{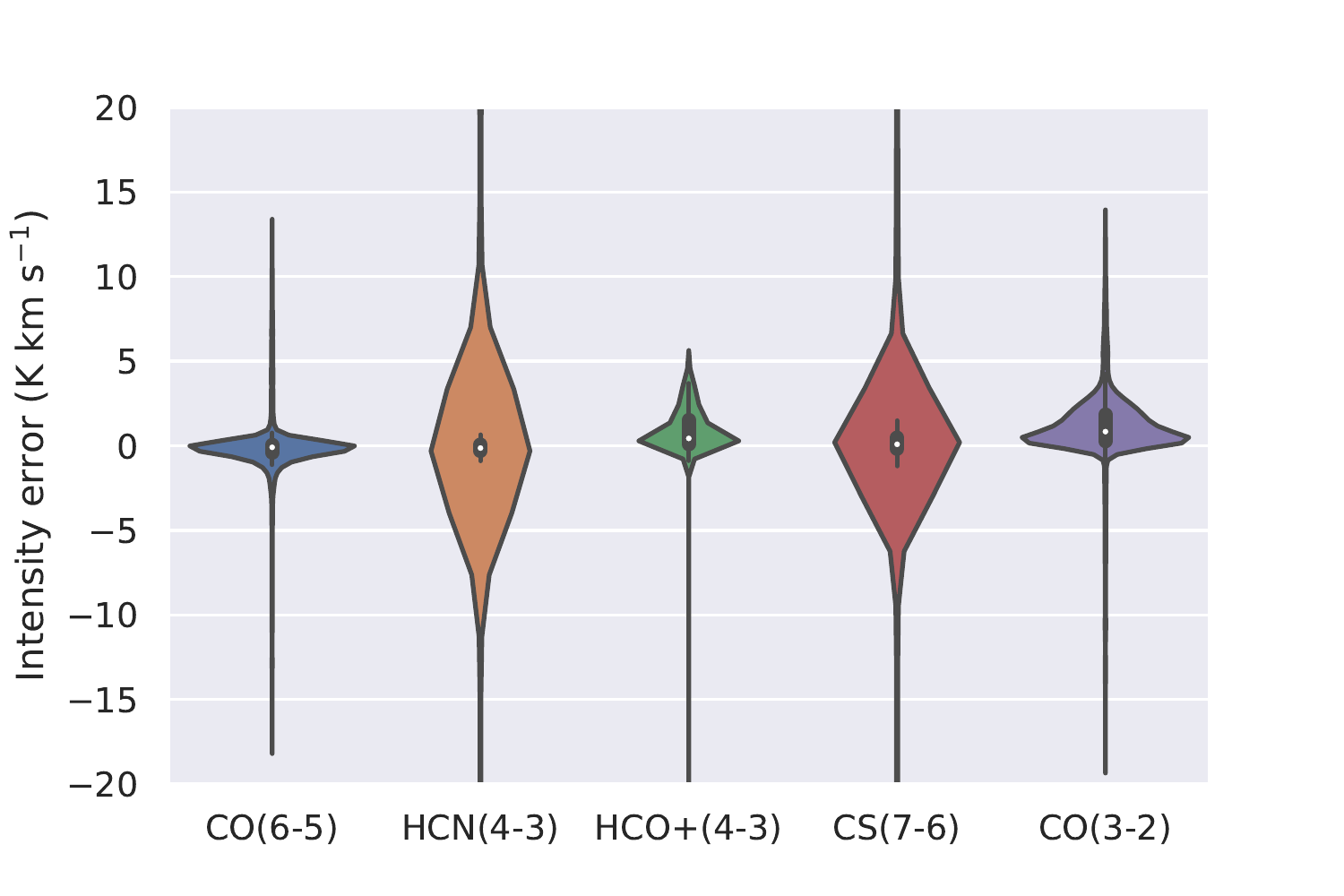}
    \caption{Violin plot of the distribution of the difference between intensity predictions from the emulator and from RADEX for different molecular lines. The distribution is obtained using a kernel density estimate from the 10000 simulations in the dataset.}
    \label{fig:violinErrorRadex}
\end{figure}

To assess our emulators effectiveness we plotted the difference between RADEX and our emulator, alongside for comparison the distribution  of RADEX intensities, for a dataset of 10000 unseen RADEX simulations for CO,CS,HCO$^+$ and HCN (Figure \ref{fig:violinIntensityRadex} and \ref{fig:violinErrorRadex}). From these figures, we can see that the errors in molecular intensity associated to using the RADEX emulator are comparatively small compared to the errors associated from using the UCLCHEM emulators.In addition, as it is unlikely that all unphysical RADEX simulations have been removed, the tails of our violin plot may be skewed from the unphysical models. From this, we see that the uncertainties associated to the RADEX emulator should be much smaller than those associated to the UCLCHEM emulator. As such, the RADEX emulator should have a minimal impact on our predictions. However, we caution that this situation would change if we were modelling multiple lines transitions as, because of the shared column density, the uncertainty from the RADEX emulator would then matter much more.

\section{Bayesian Posterior Evaluation} \label{Bayesian_retrieval}

In this paper we advocate the use of emulators as a computationally efficient way of incorporating chemical models into the estimation of model parameters. 

To assess the benefits obtained from an inclusion of a chemical model in the forward model, we contrast parameter estimation with and without chemical models. The parameter estimation was performed using Bayesian statistics. The PyMultinest implementation of the Nested Sampling algorithm (see \cite{Skilling2006}, \cite{Feroz2009} and \cite{Buchner2014}) was used for sampling from our posterior probability distributions.

In the following sections we begin by giving a brief overview of the Bayesian formalism and Nested Sampling algorithm. We then cover, using increasingly complex models, the advantages and disadvantages of the parameter estimation using our chemical emulator.

We wish to emphasize that, in the following sections, our objective is to highlight how the emulator may be used for parameter estimation. There is some level of flexibility in how the likelihood may be parametrized, and we do not claim that the parametrization we used is necessarily optimal.

\subsection{Theory}

\subsubsection{Bayesian Formalism}

Given a model governed by a set of parameters, Bayesian statistics makes it possible to mathematically quantify the effect previously unseen data has on further focusing the parameters. In this framework, a probability distribution is associated to each parameter. The prior probability distribution reflects one's belief about the parameters before accounting for data. This probability distribution will be high in regions of the parameter space likely to coincide with the true parameters and low in other regions. The posterior probability distribution represents the updated probability distribution after accounting for observations, it is mathematically related to the prior distribution through Bayes rule:

\begin{equation}
P(\theta\mid d) = \frac{P(d \mid \theta) P(\theta)}{P(d)}.
\end{equation}
In this equation, $P(\theta\mid d)$  is the posterior distribution, $P(\theta)$ the prior distribution. $P(d \mid \theta)$ is defined as the likelihood, it describes how plausible the obtained data is given a set of parameters. The denominator is called the evidence. In this project, because the data are kept constant, the evidence behaves as a normalization constant and can be ignored.

\subsection{Application}

In this section we describe how we have applied the Bayesian statistical formalism towards constraining physical parameters from observations of molecular lines. We consider the case where we have observed $N$ molecular line-transitions obtaining observations 
\begin{equation}
X=\begin{pmatrix}
x_1\\ x_2\\ ..\\x_N 
\end{pmatrix}
\end{equation}
with $x_i$ the intensity of the $i^{th}$ molecular species and we further consider that we wish to estimate  $p(\theta \mid  X)$ with $\theta$ a vector describing the molecular gas parameters of a forward model $f$. Using Bayes Rule we can express this in terms of the likelihood and the prior distribution. For our purposes, we consider that our observations correspond to the intensities predicted from our forward model with an independent additive gaussian noise. This leads to a log-likelihood of the form 
\begin{equation}
\boldmath{\textup{ln}(L(\theta))=\textup{A}-\frac{1}{2}\sum_{i=1}^{n}\frac{(x_{i}-f_i(\theta))^2}{\sigma_i^2}},
\end{equation}
with $A$ a normalization constant, $\sigma_i$ the uncertainty associated to the i$^{th}$ specie and $f(\theta)$ the vector of intensities predicted by the forward model $f$ for model parameters $\theta$. The forward model can either be "chemistry-independent" or "chemistry-dependent" (see section \ref{forward}). 

We set uniform or log-uniform priors on all the parameters with ranges matching the emulator range. The choice of uniform prior meant that all values within the prior bounds were expected to be equally likely, while the choice of log-uniform priors meant that all scales within the bounds were expected to be equally likely. Uniform priors were used for the temperature, visual extinction,  line width, filling factor and metallicity. Log-Uniform priors were used for the cosmic-ray ionization rate, uv-photoionization rate, number density and scaled column densities. Unless otherwise stated, we used parameter priors as found in Table  \ref{tab:parameters_priors}.
\begin{table}
\centering

\begin{tabular}{ p{4.2cm} p{1.8cm} p{1.6cm}}
 \hline\hline
 Parameter& prior type & range\\
 \hline
 $A_{\rm V}$ (mags) & uniform    & 1-100 \\
 n (cm$^{-3}$)&   log-uniform  & $10^4$-$10^6$   \\
 $\zeta$  ($1.3\times 10^{-17} \rm s^{-1}$)& log-uniform& $10^0$-$10^3$ \\
 $\chi$ (Draine)    & log-uniform& $10^0$-$10^3$ \\
 T (K)&   uniform  & 10-200  \\
 m$_Z$ (Z) & uniform  & 0.2-2  \\
  f (-)& uniform  & 0-1  \\
  $\Delta v$ (km s$^{-1}$)& uniform  & 1-100\\
  N(CO)/$\Delta v$ (cm$^{-2}$/(km s$^{-1}$)) & log-uniform  & $10^{13}$-$10^{19}$  \\
  N(CS)/$\Delta v$ (cm$^{-2}$/(km s$^{-1}$))& log-uniform  & $10^{10}$-$10^{18}$  \\
  N(HCN)/$\Delta v$ (cm$^{-2}$/(km s$^{-1}$))& log-uniform  & $10^{9}$-$10^{17}$  \\
  N(HCO$^+$)/$\Delta v$ (cm$^{-2}$/(km s$^{-1}$))& log-uniform  & $10^{8}$-$10^{15}$  \\

\hline
\end{tabular}
	\caption{Default prior distributions on model parameters.}
	\label{tab:parameters_priors}
\end{table}

\subsubsection{Posterior Evaluation}

It can be prohibitively expensive  to evaluate a high-dimensional posterior distribution. This motivates the use of efficient parameter-space exploration techniques which prioritize resource allocation towards exploring high-probability regions of parameter space. Because of the multimodality found in some of our posterior distributions, we used the pymultinest python module \citep{Buchner2014} to evaluate our posterior distributions, this is a python wrapper for the multinest package \citep{Feroz2009} which is itself an implementation of the nested sampling algorithm \citep{Skilling2006a}. We used the corner module \citep{corner} to visualize the marginalized posterior parameter distributions.

\subsection{One-phase model}
\subsubsection{Generation}
So as to contrast and evaluate the two forward modelling approaches, we generated a set of synthetic observations using the non-emulated UCLCHEM and RADEX. The parameters used for generating the observations can be found in Table \ref{tab:one_phase_table} and the resultant intensities can be found in Table \ref{tab:intensities_table}. In this case we assumed a known beam-filling factor of $f=1$.

\begin{table}
\centering
\begin{tabular}{p{4.5cm}p{3.5cm}}
\hline \hline
 & Model \\ \hline
\multicolumn{1}{l}{T (K)} & 150  \\ 
\multicolumn{1}{l}{n (cm$^{-3}$)}  & $5 \times 10^5$   \\ 
\multicolumn{1}{l}{m$_Z$ (Z)}  & 0.9  \\ 
\multicolumn{1}{l}{$A_{\rm V}$} (mags) & 40    \\ 
\multicolumn{1}{l}{$\chi$ (Draine)}  & 10    \\ 
\multicolumn{1}{l}{$\zeta$ ($1.3\times 10^{-17} \rm s^{-1}$) }  & 100  \\ 
\multicolumn{1}{l}{$\Delta v$ (km s$^{-1}$)}   & 50 \\ \hline
\end{tabular}
	\caption{Parameters used for creating the single-phase model.}
	\label{tab:one_phase_table}
\end{table}

\begin{table}
\centering
\begin{tabular}{p{2.0cm} p{3.0cm} p{3.0cm}}
 \hline \hline
 \multicolumn{3}{c}{Beam-Adjusted Intensities} \\
 \hline
Transition& Intensity (K km s$^{-1}$)  & Emulated Intensity (K km s$^{-1}$)\\
 \hline
CO(3-2)   &   6760.0 &   6869.1\\
CO(6-5)   &     6905.0 &   6844.5      \\
HCN(4-3)  &     905.5  &   958.3     \\
HCO$^+$(4-3) &     16.1 &   19.9      \\
CS(7-6)   &     361.05  &   443.7   \\ \hline            
\end{tabular}
	\caption{Intensities (K km s$^{-1}$) of the single-phase model.}
	\label{tab:intensities_table}
\end{table}

\subsubsection{Posterior Estimation}

\begin{figure}
	\includegraphics[width=\columnwidth]{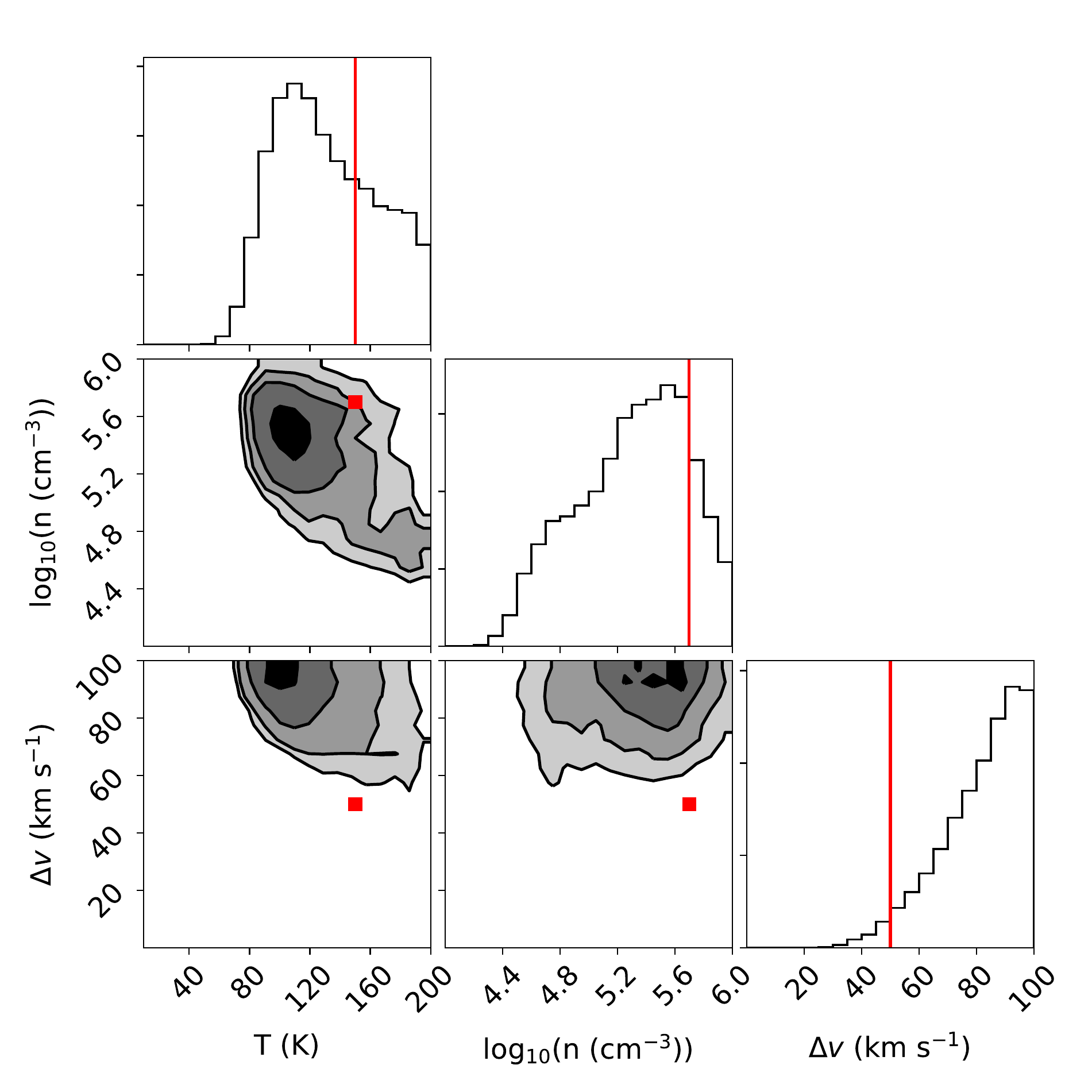}
    \caption{Marginalised posterior distributions obtained when using a single-phase "chemistry-independent" forward model. The true parameters, plotted in red, can be found in Table \ref{tab:one_phase_table}.}
    \label{fig:Plot1PhaseIntensity2Phases}
\end{figure}

\begin{figure*}
	\includegraphics[width=\linewidth]{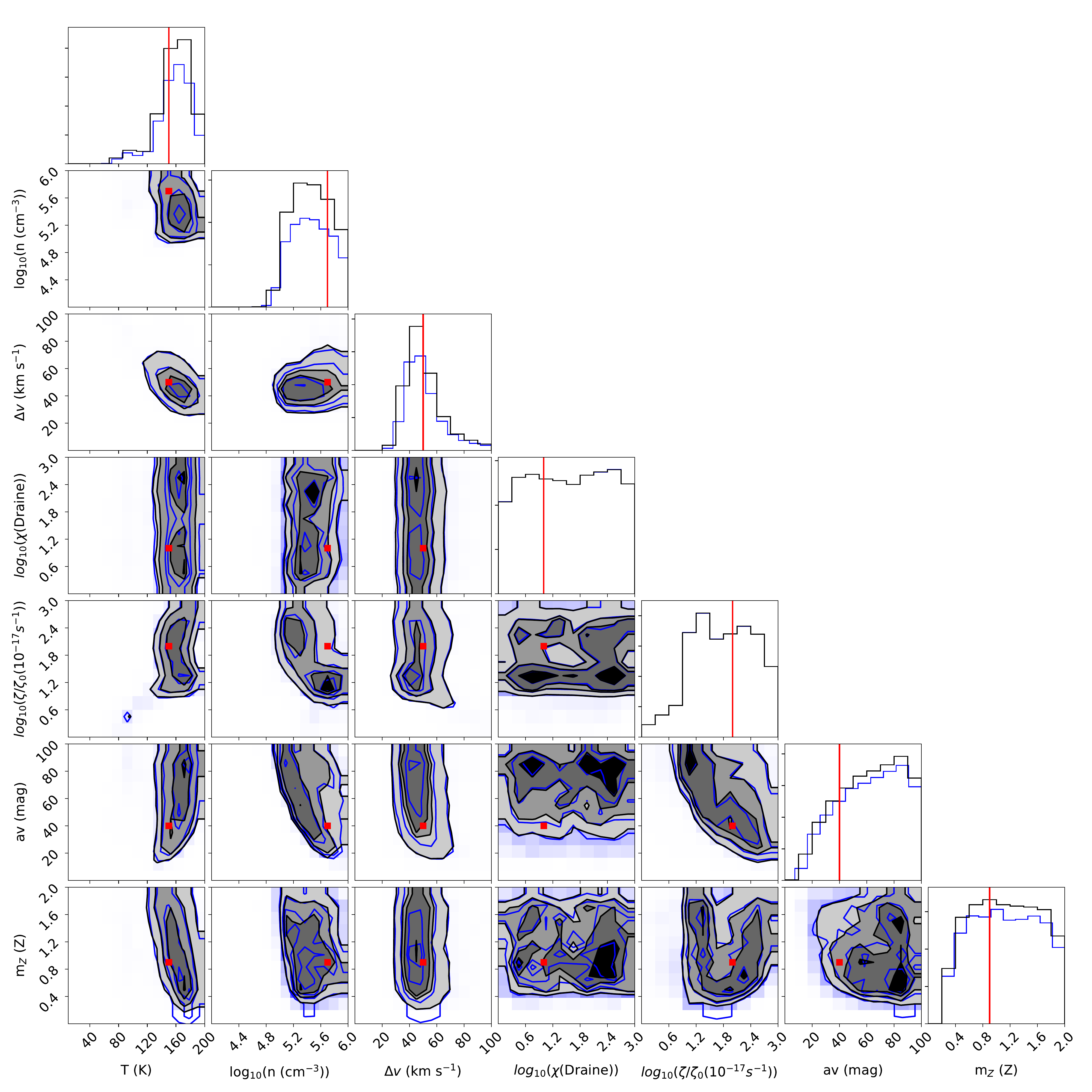}
    \caption{Marginalised posterior distributions obtained when using a single-phase "chemistry dependent" forward model. The posterior distributions obtained using the emulators are plotted in black while those obtained using the non-emulated models are plotted in blue. The true parameters, plotted in red, can be found in Table \ref{tab:one_phase_table}.}
    \label{fig:Plot1PhaseAbundanceIntensity2Phases}
\end{figure*}

We attempted to retrieve the parameters using the "chemistry dependent" forward model and the "chemistry independent" forward model. In both cases, we assumed observational uncertainties of $20\%$ and prior distributions as defined in Table \ref{tab:parameters_priors}. Marginalised one and two-dimensional posterior probability distributions for the temperature, density and linewidth, as obtained using the corner module, are shown in Figure \ref{fig:Plot1PhaseIntensity2Phases} for the "chemistry-independent" forward model and in Figure \ref{fig:Plot1PhaseAbundanceIntensity2Phases} for the "chemistry-dependent" forward model. For the "chemistry-independent" forward model, the posterior distribution was evaluated using the emulated models, while for the "chemistry-dependent" forward model, the posterior distribution was evaluated using both the emulated (black) and non-emulated model(blue).

We see that the posterior distributions obtained using the emulated and non-emulated forward models (Figure \ref{fig:Plot1PhaseAbundanceIntensity2Phases}) are in excellent agreement with each other. This further supports our findings from previous sections, that the emulator can reproduce with high fidelity the RADEX and UCLCHEM model predictions.

From these figures, it is apparent that the "chemistry-dependent" and the "chemistry-independent" parameter estimation give very different predictions. While the "chemistry-independent" parameter estimation struggles to constrain the temperature and density of the molecular gas, the "chemistry-dependent" estimation is able to return tight and accurate confidence bounds on the parameters.

\subsection{Two-phase model}
To further test our parameter retrieval process, we modeled a new molecular gas phase by generating an additional set of intensities, approximating a lower temperature molecular phase, using the non-emulated UCLCHEM and RADEX. We then modelled a beam containing two molecular gas phases by adding the two phases after scaling them by beam-filling factors. The parameters used for generating the two phases and the intensities of the two phases can be found in Tables \ref{tab:parameters_table} and \ref{tab:intensities_table_two}. These parameters roughly correspond  to a beam filled with one hot and dense phase and one cool and diffuse phase.

\begin{table}
	\centering

\begin{tabular}{p{2.1cm} p{3.0cm} p{2.9cm}}
\hline \hline
                           & Model I & Model II \\ \hline
\multicolumn{1}{l}{T (K)}              & 50             & 150               \\ 
\multicolumn{1}{l}{n (cm$^{-3}$)}              & $2\times 10^4$   & $5 \times 10^5$   \\ 
\multicolumn{1}{l}{m$_Z$ (Z)}    & 0.9              & 0.9               \\
\multicolumn{1}{l}{$A_{\rm V}$} (mags)             & 3               & 40                \\ 
\multicolumn{1}{l}{$\chi$ (Draine)}       & 10               & 10                \\ 
\multicolumn{1}{l}{$\zeta$ ($1.3\times 10^{-17} $s$^{-1}$)}           & 10               & 100               \\
\multicolumn{1}{l}{f (-)} & 0.7              & 0.3              \\ 
\multicolumn{1}{l}{$\Delta v$ (km s$^{-1}$)}     & 50               & 50            \\ \hline
\end{tabular}
	\caption{Parameters used for creating the two-phase model.}
	\label{tab:parameters_table}
\end{table}

\begin{table}
\centering
\begin{tabular}{p{1.9cm} p{3.0cm} p{3.0cm}}
 \hline \hline
 \multicolumn{3}{c}{Beam-Adjusted Intensities} \\
 \hline
Transition& Intensity (K km s$^{-1}$)  & Emulated Intensity (K km s$^{-1}$)\\
 \hline
CO(3-2)   &   2865.2 &   2930.0\\
CO(6-5)   &     2233.4 &   2225.7      \\
HCN(4-3)  &     271.9  &   287.5     \\
HCO$^+$(4-3) &     4.9 &   6.9      \\
CS(7-6)   &     108.5  &   124.0   \\ \hline            
\end{tabular}
	\caption{Intensities (K km s$^{-1}$) of the two-phase model.}
	\label{tab:intensities_table_two}
\end{table}

\subsubsection{Two-phase parameter estimation}

Much like in the previous section, we next attempted to fit our two-phase observations using "chemistry-dependent" and "chemistry-independent" forward models. In both cases, we used the emulated two-phase forward models with prior distributions identical to those used in the single-phase forward models (Table \ref{tab:parameters_priors}) on both phases and assumed uncertainties of $20\%$ on the observations \citep{Viti2014}.  As we set both prior distributions to cover the full emulated-range, the forward model could hypothetically fit the observations with two hot gas phases or two cold gas phases. This is in contrast with what has sometimes been done, such as in \cite{Tunnard2015}, where they have constrained the two phases to non-overlapping parameter ranges thus artificially enforcing the gas to exist in two very distinct phases. In this analysis we have forced the two phases of gas to share the same metallicity and line-width.

\begin{figure}
	\includegraphics[width=\columnwidth]{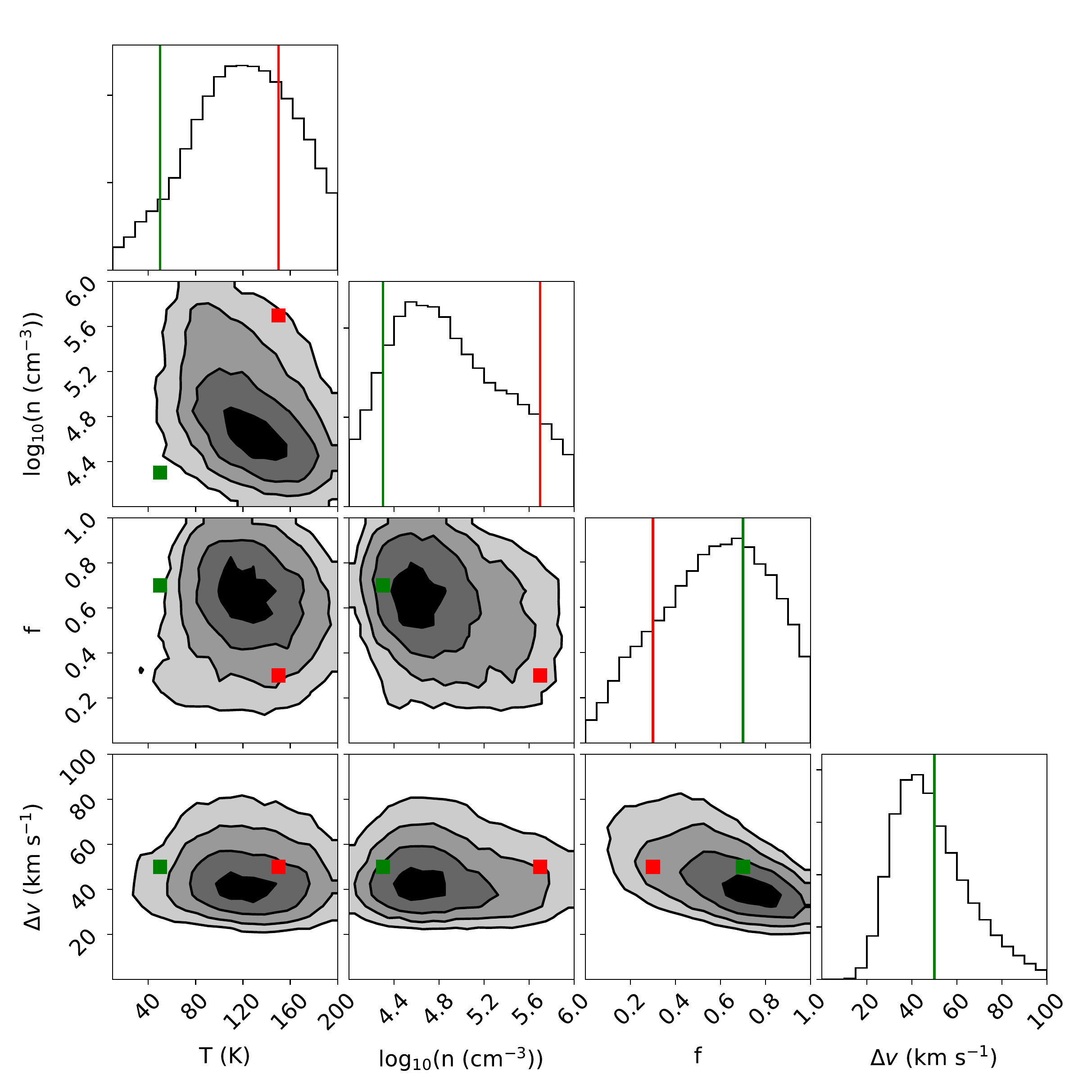}
    \caption{Marginalised posterior distributions obtained when using a two-phase "chemistry-independent" forward model. The true parameters, plotted in green and red, can be found in Table \ref{tab:parameters_table}.}
    \label{fig:Plot2PhaseIntensity2Phases}
\end{figure}

\begin{figure*}
	\includegraphics[width=\linewidth]{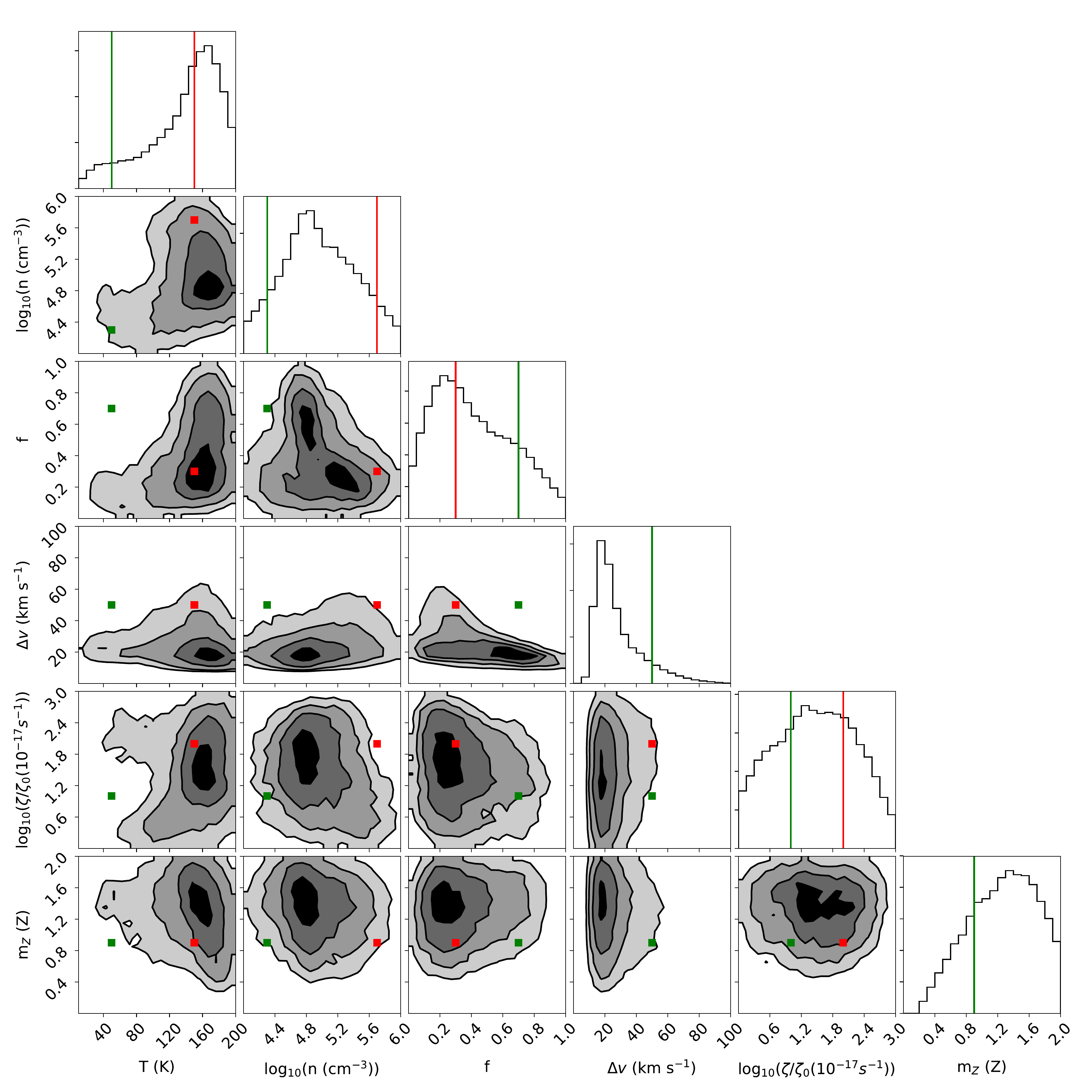}
    \caption{Marginalised posterior distributions obtained when using a two-phase "chemistry-dependent" forward model. The true parameters, plotted in green and red, can be found in Table \ref{tab:parameters_table}.} 
    \label{fig:Plot2PhaseAbundanceIntensity2Phases}
\end{figure*}

The marginalised posterior distributions obtained are plotted in Figure \ref{fig:Plot2PhaseIntensity2Phases} and Figure \ref{fig:Plot2PhaseAbundanceIntensity2Phases}. Once again, we see that both approaches give very different constraints on the posterior probability distributions. Although, both methods arguably struggle to retrieve the correct temperature and density, the "chemistry-dependent" forward model predictions occupy a more narrow range of the parameter-space. 

In addition, the "chemistry-dependent" forward model also partially recovers the two-phase bimodality. Even though this is most apparent in the plot of filling factor against density, it is also visible in the temperature vs density plot. All of this is indicative that the chemistry-dependent estimation is capable, at least partially, of picking-out the two distinct phases of gas, even when the chemistry-independent forward model struggles to constrain any parameter. On the other hand, we see that our parameter estimation underestimates the linewidth.

\section{Application to real line ratios} \label{Applications_Real}

In the previous section, we have used synthetic observations to assess the benefits brought by incorporating chemistry into radiative-transfer forward models. However, although synthetic observations are useful in that they offer a controllable and well understood testbed, there are aspects of working with real regions that cannot be easily understood with synthetic observations.

Indeed, a non-comprehensive list of these complications are:
\begin{itemize}
  \item Even though in recent years there has been tremendous progress towards understanding the chemistry in the interstellar medium \citep{williams_viti_2013}, there are still significant uncertainties associated to the reactions in the interstellar medium. Because of these, the chemistry in our forward model may not accurately match that occurring in real regions.
  \item Our emulator is only usable/valid for the parameter range under which it was trained (Table \ref{tab:parameters_introduction}).
  \item The molecular abundances predicted by our emulator are those reached after the chemical models have been run long enough for an equilibrium to be reached. As such any transient chemical variation will not be captured by our emulator.
  \item For molecular observations, and particularly for observations with large beam-sizes, the approximation that the gas can be represented using a small number of components is likely to break down. 
\end{itemize}

For the reasons highlighted above, we though it important to showcase the performance of the emulator on real observations. To do this, we used ALMA observations of NGC1068 a prototypical nearby Seyfert barred galaxy, as presented in  \cite{Viti2014} and \cite{Viti2017}. This galaxy is believed to host a rich chemical diversity, as such it is expected to be very difficult to disentangle the chemistry occurring within. We focused on the spectral lines measured in the East Knot, a region of the molecular ring showing strong emission, and used the degraded resolutions measurements as  presented in the original paper \citep{Viti2014}, the measured intensities have been retranscribed and can be found in Table \ref{tableIntensities}. Although we have every reason to believe that the molecular gas spans a wide range of physical conditions, to avoid using an excessively complex forward model, we fit the region using a single-phase molecular component.

After running an exploratory single-phase "chemistry-dependent" parameter estimation, we found that our models were consistently unable to reproduce the HCO$^+$ intensities. To further investigate this issue, we evaluated the HCO$^+$(4-3)/CO(3-2) line-intensity ratio for all of the UCLCHEM models used by our emulator. We found that all of the chemical models in our training dataset resulted in line-intensity ratios smaller than the ratio observed with ALMA. This suggests that our models, over the parameter-range studied, does not produce enough HCO$+$ to match the observations.

This was in itself not surprising, as chemical models have been known to struggle with producing enough HCO$^+$ to match observations. Some examples of this can be found in \cite{Godard2010} and \cite{Viti2014}. In \cite{Papadopoulos2007}, it was argued that because of the sensitivity of the HCO$^+$ column density to the ionization degree of the molecular gas, HCO$^+$ can be an unreliable tracer of hot dense gas. Furthermore, if the HCO$^+$ is a transient species or if it is created in low visual extinction and high density clumps not covered by our emulation, then it is likely that our models would not capture it.

\begin{figure*}
	\includegraphics[width=\linewidth]{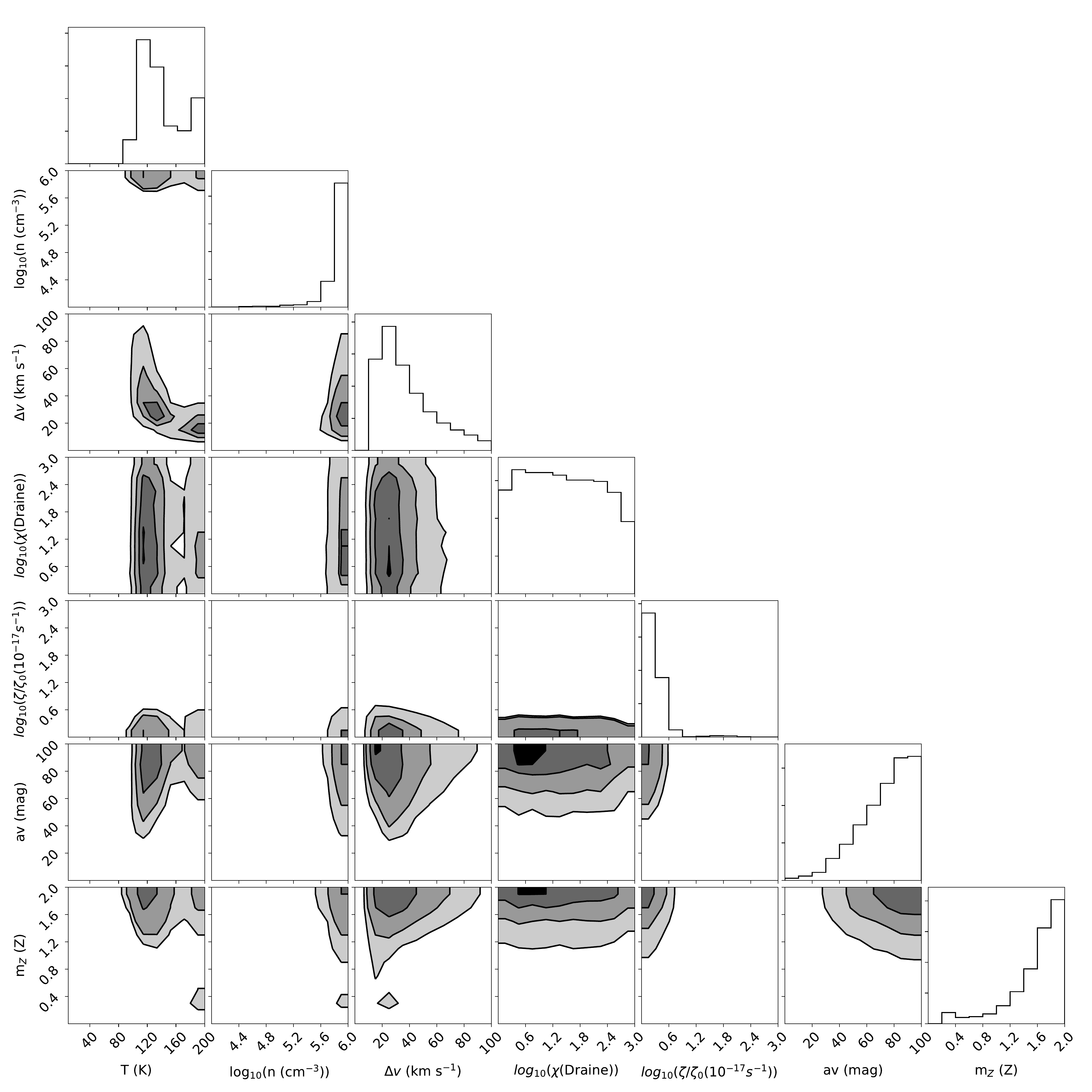}
    \caption{Marginalised posterior distributions obtained when using a single-phase "chemistry-dependent" forward model on the ALMA observations excluding HCO$^+$.}
    \label{fig:Plot2PhaseAbundanceIntensityRealAlmaNoHCO+}
\end{figure*}

In light of our inability to reproduce HCO$^+$ observations, we reran a parameter estimation identical in all but the fact that HCO$^+$ was excluded from the fitting.  The posterior plots obtained without HCO$^+$ can be found in Figure \ref{fig:Plot2PhaseAbundanceIntensityRealAlmaNoHCO+}. From this we can see that the Bayesian parameter estimation, for fitting a single phase, favours a component with moderately high temperature ($T \sim 120 K$) but very high density ($n \sim 10^6 cm^{-3}$).

It is informative to quantify the goodness of fits of some of the models from the posterior distributions. We show, for a small representative sample of well-fit models, the model parameters in Table \ref{tableModels} and the corresponding intensities in Table \ref{tableIntensities}. From these Tables, it becomes clear that the intensities predicted by the emulated and non-emulated forward model are in excellent agreement. These tables also highlight the strong degeneracies which exist in the forward modelling process. 

One thing of note is that the HCN intensity recovered by the Bayesian parameter estimation, although at the correct order of magnitude, was consistently lower than the observed HCN intensity. Almost none of the best-fitting models predicted an HCN intensity greater than the observed intensity. This could be interpreted as the models struggling to create a high enough HCN intensity which could be indicative that the molecular phase not captured by our models which is responsible for the high HCO$^+$ intensity may also be at least partially responsible for a portion of the HCN intensity.

\begin{table*}
\centering
\begin{tabular}{p{0.06\linewidth}p{0.08\linewidth}p{0.10\linewidth}p{0.10\linewidth}p{0.10\linewidth}p{0.08\linewidth}p{0.08\linewidth}p{0.10\linewidth}p{0.06\linewidth}}
\hline \hline
model & $\chi$ (Draine) & $\zeta$  ($1.3\times 10^{-17} \rm$) & n (cm$^{-3}$)  & $A_{\rm V}$ (mags)  & T (K) & m$_Z$ (Z) & $\Delta v$ (km s$^{-1}$)& f (-) \\
\hline
(1)&237.90&1.24&966810.74&89.10&199.24&0.26&18.81&0.75 \\
\hline
(2)&90.46&1.02&979300.34&86.86&108.46&1.30&78.15&0.31 \\
\hline
(3)&31.28&2.95&907548.96&71.29&102.80&1.90&99.09&0.27 \\
\hline
(4)&13.94&1.39&857614.52&47.84&96.72&1.85&58.51&0.44  \\
\end{tabular}
\caption{Example input model parameters. For the associated intensities see Table \ref{tableIntensities}.}
\label{tableModels}
\end{table*}

\begin{table*}
\centering
\begin{tabular}{p{0.1\textwidth}p{0.1\textwidth}p{0.1\textwidth}p{0.1\textwidth}p{0.1\textwidth}p{0.1\textwidth}}
\hline \hline
model & CO(3-2) & CO(6-5) & HCN(4-3)  & HCO$^+$(4-3)  & CS(7-6)\\
\hline 
(1) emul& 2397.37&2656.13&532.01&8.38&8.89\\
(1) direct& 2566.15&2705.82&485.44&5.95&9.07\\
\hline 
(2) emul& 2611.54&2347.36&470.32&1.73&7.86\\
(2) direct& 2577.70&2377.35&554.30&0.94&3.96\\
\hline
(3) emul& 2710.83&2426.85&451.47&1.55&10.30\\
(3) direct& 2688.81&2465.41&593.14&1.13&11.54\\
\hline
(4) emul& 2491.62&2215.54&454.70&1.47&9.97\\
(4) direct& 2422.29&2209.38&656.22&0.72&5.33\\
\hline
\hline
\textbf{observed}&\textbf{2346.28}&\textbf{2712.70}&\textbf{639.47}&\textbf{251.18}&\textbf{8.26}\\
\end{tabular}
\caption{Intensities (in K km s$^{-1}$) obtained for the models as defined in Table \ref{tableModels}. The emul columns correspond to the intensities obtained using the emulated UCLCHEM and emulated RADEX. The direct columns correspond to the intensities obtained from using the true UCLCHEM and true RADEX. The last column contains the measured NGC1068 intensities for comparison.}
\label{tableIntensities}
\end{table*}

\section{Conclusions}

In this paper we have proposed an alternative method for interpreting the physical conditions of the interstellar medium from the analysis of molecular line intensities. This is typically approached by running many forward models and finding the input parameters to the non-LTE radiative-transfer forward-model whose predictions match well with observations.
 
By feeding the outputs of chemical model as inputs to the RADEX radiative transfer model, as was expanded upon in section \ref{forward}, it is possible to re-parametrize the radiative-transfer forward model. In this formalism, the forward-model dependency on column-densities is replaced with a dependency on physical parameters. This offers the potential of lifting parameter degeneracies. However, the subsequent forward model is not computationally practical as the required chemical models are computationally costly to run at scale.

We presented and evaluated an emulator, created using neural networks, capable of predicting molecular abundances at a fraction of the run-time of the UCLCHEM astrochemical model as well as an emulator approximating the outputs of the RADEX radiative transfer codes. We show, that by using our emulators, it becomes computationally tractable to run the chemistry-dependent forward models accurately at scale.

By applying our emulator to mock observations as well as real observations we have shown that incorporating chemistry into the parameter retrieval can not only lead to tighter constraints on the retrieved physical parameters but also constrain parameters introduced by the chemical models such as the cosmic-ray ionization rate and the metallicity. We also showed that  our emulator-based approach was able to distinguish between two distinct phases of molecular gas where a traditional radiative transfer approach failed.

Finally, by applying our formalism to real observations of the galaxy NGC1068, we showed that our emulator can effectively be applied to get insights about real observations. This comes with the caveat that the emulator may struggle to reproduce certain molecular lines such as the HCO$^+$ molecular lines. However, we argued that this inability of reproducing HCO$^+$ may be indicative of molecular gas with extreme physical conditions not within our emulation range. 

We would like to conclude this paper by emphasising that we have had to make choices in defining our likelihood for our experiments, but that these choices may not be the best choices. For example, the likelihood could be designed to not only put emphasis on having the line intensities be at the correct scale, as was done in our experiments, but also put emphasise on preserving the relative strength of lines tracing the same specie. Finally, the likelihood could also be designed to put stronger constraints on species for which the chemical modelling has been benchmarked and well understood. Furthermore, in most applications it is probably sensible to further constrain or fix some of the forward model parameters, such as the metallicity.

\begin{acknowledgements}
DDM is funded by an STFC studentship in data-intensive science. The authors also acknowledge DiRAC for use of their HPC system which allowed this work to be performed.
\end{acknowledgements}

\bibliographystyle{aa} 
\bibliography{emulator} 
\end{document}